\documentclass[conference]{IEEEtran}
\IEEEoverridecommandlockouts
\usepackage{cite}
\usepackage{amsmath,amssymb,amsfonts}
\usepackage{algorithmic}
\usepackage{graphicx}
\usepackage{textcomp}
\usepackage{tikz}
\usepackage{pgfplots}
\usepackage{tabu}
\usepackage{colortbl}
\usepackage{array}
\usepackage{caption}
\usepackage{subcaption}
\usepackage{float}
\usetikzlibrary{shapes}
\usetikzlibrary{shapes,arrows}
\usetikzlibrary{fit}
\usetikzlibrary{shapes.multipart}
\usetikzlibrary{positioning}
\usetikzlibrary{shapes,backgrounds,calc}
\usetikzlibrary{calc,shapes.geometric}
\usepackage{float}

\usetikzlibrary{pgfplots.groupplots}
\usetikzlibrary{patterns}
\usepackage[linesnumbered,ruled,noend]{algorithm2e}
\usetikzlibrary{fit,shapes}

\definecolor{darkgreen}{rgb}{0,0.65,0}%
\usetikzlibrary{arrows.meta}
\usetikzlibrary{arrows}

\newcounter{lemma}

\newtheorem{exampleplain}{Example}

\newenvironment{example}{\begin{exampleplain}}{~\hfill$\vartriangle$\end{exampleplain}}

\begin{document}
%
\title{Decoding Staircase Codes with Marked Bits
\thanks{This work is in part supported by the NSFC Program (No. 61431003, 61601049, 61625104, and 61701155) and Fund of State Key Laboratory of Information Photonics and Optical Communications, Beijing University of Posts and Telecommunications (No. IPOC2017ZT08). The work of A. Alvarado is supported by the Netherlands Organisation for Scientific Research (NWO) via the VIDI Grant ICONIC (project number 15685) and has received funding from the European Research Council (ERC) under the European Union's Horizon 2020 research and innovation programme (grant agreement No 57791). The author Yi Lei would like to thank China Scholarship Council (CSC) for supporting her study in Netherlands.}
}

\author{\IEEEauthorblockN{Yi Lei\IEEEauthorrefmark{1}\IEEEauthorrefmark{2},
Alex Alvarado\IEEEauthorrefmark{2},
Bin Chen\IEEEauthorrefmark{2}\IEEEauthorrefmark{3},
Xiong Deng\IEEEauthorrefmark{2},
Zizheng Cao\IEEEauthorrefmark{2},
Jianqiang Li\IEEEauthorrefmark{1} and
Kun Xu\IEEEauthorrefmark{1}}
\IEEEauthorblockA{\IEEEauthorrefmark{1}State Key Laboratory of Information of Photonics and Optical Communications, \\
Beijing University of Posts and Telecommunications (BUPT),China}
\IEEEauthorblockA{\IEEEauthorrefmark{2}Department of Electrical Engineering, Eindhoven University of Technology (TU/e), The Netherlands}
\IEEEauthorblockA{\IEEEauthorrefmark{3}School of Computer and Information, Hefei University of Technology (HFUT), Hefei, China\\
Emails: leiyi@bupt.edu.cn, a.alvarado@tue.nl}}

\maketitle

\begin{abstract}
Staircase codes (SCCs) are typically decoded using iterative bounded-distance decoding (BDD) and hard decisions. In this paper, a novel decoding algorithm is proposed, which partially uses soft information from the channel. The proposed algorithm is based on marking certain number of highly reliable and highly unreliable bits. These marked bits are used to improve the miscorrection-detection capability of the SCC decoder and the error-correcting capability of BDD. For SCCs with $2$-error-correcting BCH component codes, our algorithm improves upon standard SCC decoding by up to $0.30$~dB at a bit-error rate of $10^{-7}$. The proposed algorithm is shown to achieve almost half of the gain achievable by an idealized decoder with this structure. 
\end{abstract}

\IEEEpeerreviewmaketitle

\section{Introduction}

Forward error correction (FEC) is required in optical communication systems to meet the ever increasing data demands in optical transport networks (OTNs), currently targeting data rates of $400$~Gb/s and beyond\cite{400G1,400G3}. As the data rates increase, FEC codes that can boost the net coding gain (NCG) are of key importance. Soft-decision FEC codes provide large NCGs, however, they are not the best candidates for very high data rate applications due to their high power consumption and decoding delay. In this context, simple but powerful hard-decision (HD) FEC codes are a promising alternative, e.g., Reed-Solomon (RS) code \cite{G975} and concatenated codes consisting of two HD codes\cite{G9751}. 
One popular family of HD-FEC codes is the so-called staircase codes (SCCs) \cite{Smith2012,Zhang2014}. Compared to the best code from ITU-T standards\cite{G975,G9751}, SCCs offer an improvement of $0.42$~dB NCG \cite{Smith2012}. An implementation agreement has been reached for using an SCC as an outer code in the baseline draft of $400$~Gb/s OTN \cite{400G5}.

Similar to classical product codes, SCCs are based on simple component codes, Bose-Chaudhuri-Hocquenghem (BCH) codes being the most popular ones. SCC decoding is done iteratively based on bounded-distance decoding (BDD) for the component codes. Although very simple, one drawback of BDD is that its error-correcting capability is limited to $t=\lfloor\frac{d_{0}-1}{2}\rfloor$, where $d_{0}$ is the minimum Hamming distance (MHD) of the component code\cite{BDD}. 
BDD can detect more than $t$ errors, but cannot correct them. In some cases, BDD may also erroneously decode a received sequence with more than $t$ errors, a situation known as a \emph{miscorrection}. Miscorrections are known to degrade the performance of iterative BDD. To prevent miscorrections, several methods have been studied in the literature \cite{SmithPhD,ChristianJournal,Christian1,Alireza}.

The authors of \cite{SmithPhD} proposed rejecting bit-flips from the decoding of bit sequences associated with the last SCC block if they conflict with a zero-syndrome codeword from the previous block. However, the obtained gains are expected to be limited \cite[Sec.~I]{ChristianJournal}. An anchor-based decoding algorithm has been proposed in \cite{ChristianJournal,Christian1}, where some bit sequences are labeled as anchor codewords. These sequences are thought to have been decoded without miscorrections. Decoding results that are inconsistent with anchor codewords are discarded. 
The algorithm in \cite{Christian1} outperforms \cite{SmithPhD}, but it suffers from an increased complexity as anchor codewords need to be tracked during iterative BDD. Very recently, a modified iterative BDD for product codes was proposed in \cite{Alireza}. In this algorithm, channel reliabilities are used to perform the final HD at the output of BDD, instead of directly accepting the decoding result. Large gains are obtained, but it requires additional memory (and processing) as all the soft information needs to be saved. Moreover, its effectiveness for SCCs has not yet been reported in the literature.

In this paper, we propose a simple algorithm to improve the decoding of SCCs. This is achieved by marking highly reliable and highly unreliable bits. Unlike previous works, our proposed algorithm \emph{jointly} increases the miscorrection-detection capability of the SCC decoder and the error-correcting capability of BDD. The proposed algorithm only requires modifications to the decoding structure related to the last block of each decoding window. Furthermore, the algorithm is based on marking bits only, and thus, no soft bits (log-likelihood ratios, LLRs) need to be  saved. Marked bits do not need to be tracked during the iterative process either.


\section{System Model, SCCs, and BDD}

\subsection{System Model}

\begin{figure}[t]
\centering
\includegraphics[width=0.5\textwidth]{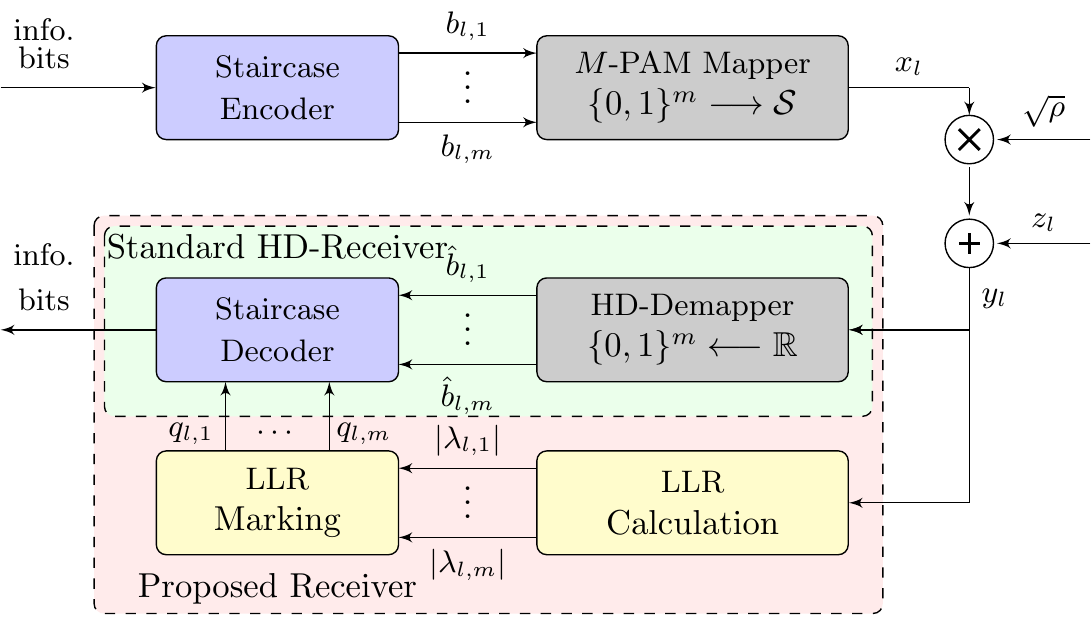}
\caption{System model under consideration.}
\vspace{-2.1ex}
\label{fig: System Model}
\end{figure}

As shown in Fig. \ref{fig: System Model}, information bits are encoded by a staircase encoder and then mapped to symbols $x_{l}$ taken from an equally-spaced $M$-ary PAM constellation $\mathcal{S}=\{s_{1},s_{2},\ldots,s_{M}\}$ with $M=2^m$ points, where $l$ is the discrete time index. The bit-to-symbol mapping is the binary reflected Gray code. The received signal is ${y_{l}}=\sqrt{\rho}{x_{l}}+{z_{l}}$, where ${z_{l}}$ is zero-mean unit-variance additive white Gaussian noise. 

The standard HD receiver structure for SCCs uses an HD-based demapper to estimate the code bits, which are then fed to the decoder (green block in Fig.~\ref{fig: System Model}). In this paper, we introduce a novel receiver architecture where the HD-FEC decoder uses partial soft information from the channel. This soft information is typically represented using LLRs, calculated as \cite[eq.~(3.50)]{LLR}
\begin{equation}\label{LLR}
  \lambda_{l,k}=\sum_{b \in \{0,1\}} (-1)^{\bar{b}} \log\sum_{i \in \mathcal{I}_{k,b}} \textrm{exp}\left(-\frac{(y_{l}-\sqrt{\rho}s_{i})^{2}}{2}\right),
\end{equation}
with $k=1,\ldots,m$, and where $\bar{b}$ denotes bit negation. In \eqref{LLR}, the set $\mathcal{I}_{k,b}$ enumerates all the constellation points in $\mathcal{S}$ whose $k$th bit $c_{i,k}$ is $b$, i.e., $\mathcal{I}_{k,b}\triangleq \{i=1,2,\ldots,M: c_{i,k}=b\}$.

Our proposed structure is shown in Fig.~\ref{fig: System Model} (red block). In this structure, apart from the HD-estimated bits $\hat{b}_{l,1},\ldots,\hat{b}_{l,m}$, a sequence of \emph{marked} bits will also be made available to the decoder. These marked bits are denoted by $q_{l,k}$ and can be: highly reliable bits (HRBs), highly unreliable bits (HUBs), or neither. The marking is made based on the absolute value of the LLRs $|\lambda_{l,k}|$. More details about the marking procedure and how this can be exploited by the staircase decoder will be given in Sec.~\ref{sec:algorithm}.

\subsection{Staircase Codes}

\begin{figure}
\centering
\includegraphics[width=0.35\textwidth]{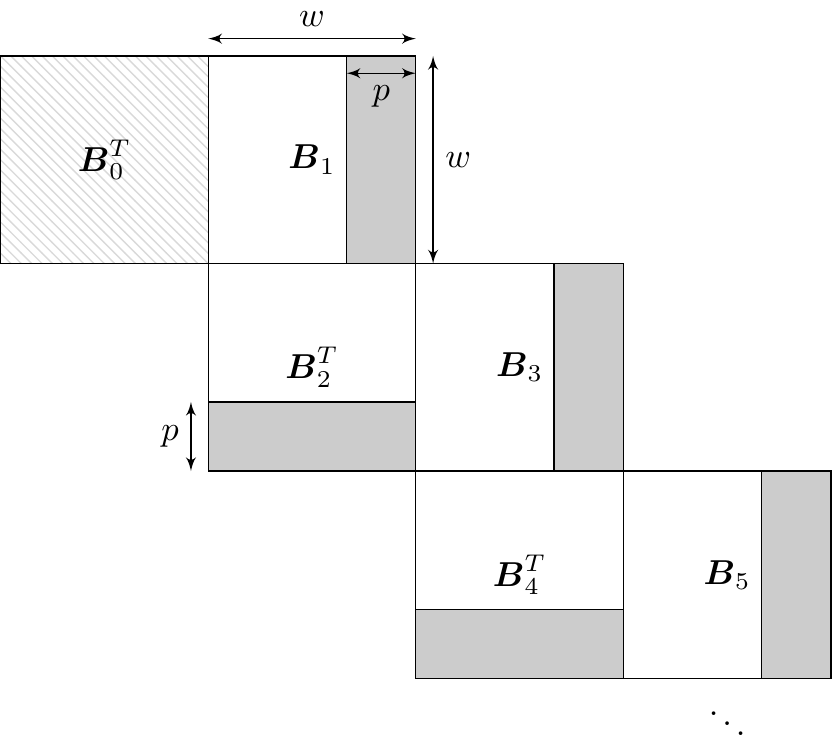}
  \captionof{figure}{Staircase structure of SCCs considered in this paper.}
  \vspace{-2ex}
  \label{fig: Structure of SCC}
\end{figure}

Fig. \ref{fig: Structure of SCC} shows the staircase structure of SCCs we consider in this paper, where $\boldsymbol{B}_{0}$ is initialized to all zeros. Each subsequent SCC block $\boldsymbol{B}_{i},i=1,2,\ldots$, is composed of $w(w-p)$ information bits (white area) and $wp$ parity bits (gray area). Each row of the matrix $[\boldsymbol{B}^{T}_{i-1} \boldsymbol{B}_{i}]$ $\forall i>1$ is a valid codeword in a component code $\mathcal{C}$. 

We consider the component code $\mathcal{C}$ to be a binary code with parameters $(n_{c}, k_{c}, t)$, where $n_{c}$ is the codeword length and $k_{c}$ is the information length. Then, $w$ and $p$ are given by: $w=n_{c}/2$ and $p=n_{c}-k_{c}$. The code rate $R$ of the SCC is $R=1-p/w=2k_{c}/n_{c}-1$. Throughout this paper, the component codes $\mathcal{C}$ considered are extended (by 1 additional parity bit) BCH codes. The mapping between code bits and symbols is done by reading row-by-row the blocks $\boldsymbol{B}_{i},i=1,2,\ldots$

At the receiver side, SCCs are decoded iteratively using a sliding window covering $L$ blocks. We use $\boldsymbol{Y}_{i}$ to indicate the received SCC block after HD-demapping corresponding to the transmitted block $\boldsymbol{B}_{i}$. The decoder first iteratively decodes the blocks $\{\boldsymbol{Y}_{0},\boldsymbol{Y}_{1},\ldots, \boldsymbol{Y}_{L-1}\}$. When a maximum number of iterations is reached, the decoding window outputs the block $\boldsymbol{Y}_{0}$ and moves to decode the blocks $\{\boldsymbol{Y}_{1}, \boldsymbol{Y}_{2},\ldots, \boldsymbol{Y}_{L}\}$. The block $\boldsymbol{Y}_{1}$ is then delivered and operation continues on $\{\boldsymbol{Y}_{2}, \boldsymbol{Y}_{3},\ldots, \boldsymbol{Y}_{L+1}\}$. This process continues indefinitely. Multiple decoding scheduling alternatives exist (see, e.g., \cite[Sec.~IV]{Smith2012}\cite[Sec.~II]{Zhang2014}). We chose the most popular one, namely, alternated decoding of pairs of SCC blocks within a window, from the bottom right to the top left of the SCC window.

\subsection{Bounded-Distance Decoding}

BDD is used to decode (in Hamming space) the received bit sequence for the component code $\mathcal{C}$. To correct up to $t$ errors, the MHD $d_{0}$ of $\mathcal{C}$ must satisfy $d_{0}\geq 2t+1$ ($d_{0}\geq 2t+2$ for extended BCH codes). Thus, every codeword in the code $\mathcal{C}$ can be associated to a sphere of radius $t$. Within such a sphere, no other codewords exist. If the received sequence $r$ falls inside one of these spheres, BDD will decode $r$ to the corresponding codeword. Otherwise, BDD will declare a failure. For a given transmitted codeword $c$ and a received sequence $r$, the BDD output $\hat{c}$ is thus given by
\begin{equation}\label{BDDequation}
    \begin{aligned}
    \hat{c}&= \left\{
    \begin{array}{lcl}
    c,  &      & \textrm{if~} d_{\textrm{H}}(r,c) \leq t,  \\
   \tilde{c} \in \mathcal{C},  &      & \textrm{if~} d_{\textrm{H}}(r,c) > t \textrm{~and~} d_{\textrm{H}}(r,\tilde{c}) \leq t,  \\
    r, &      & \textrm{if~} d_{\textrm{H}}(r,\tilde{c}) > t~\forall \tilde{c} \in \mathcal{C}. \\
    \end{array}
    \right.
    \end{aligned}
\end{equation}
where $d_{\textrm{H}}(\cdot,\cdot)$ represents the Hamming distance. In practice, BDD is often a syndrome-based decoder that uses syndromes to estimate the error pattern $e$. If the syndromes are all zeros, no errors are present. For the first two cases in \eqref{BDDequation}, BDD will declare decoding success and $\hat{c}=r\oplus e$. In the second case, although BDD will still return an error pattern $e$, this case corresponds to  a miscorrection. In the next section we will show how to improve miscorrection detection (MD) using the underlying structure of SCCs and the marked HRBs.

\section{The Proposed Algorithm}\label{sec:algorithm}

\begin{figure}[tb]
\centering
\includegraphics[width=0.5\textwidth]{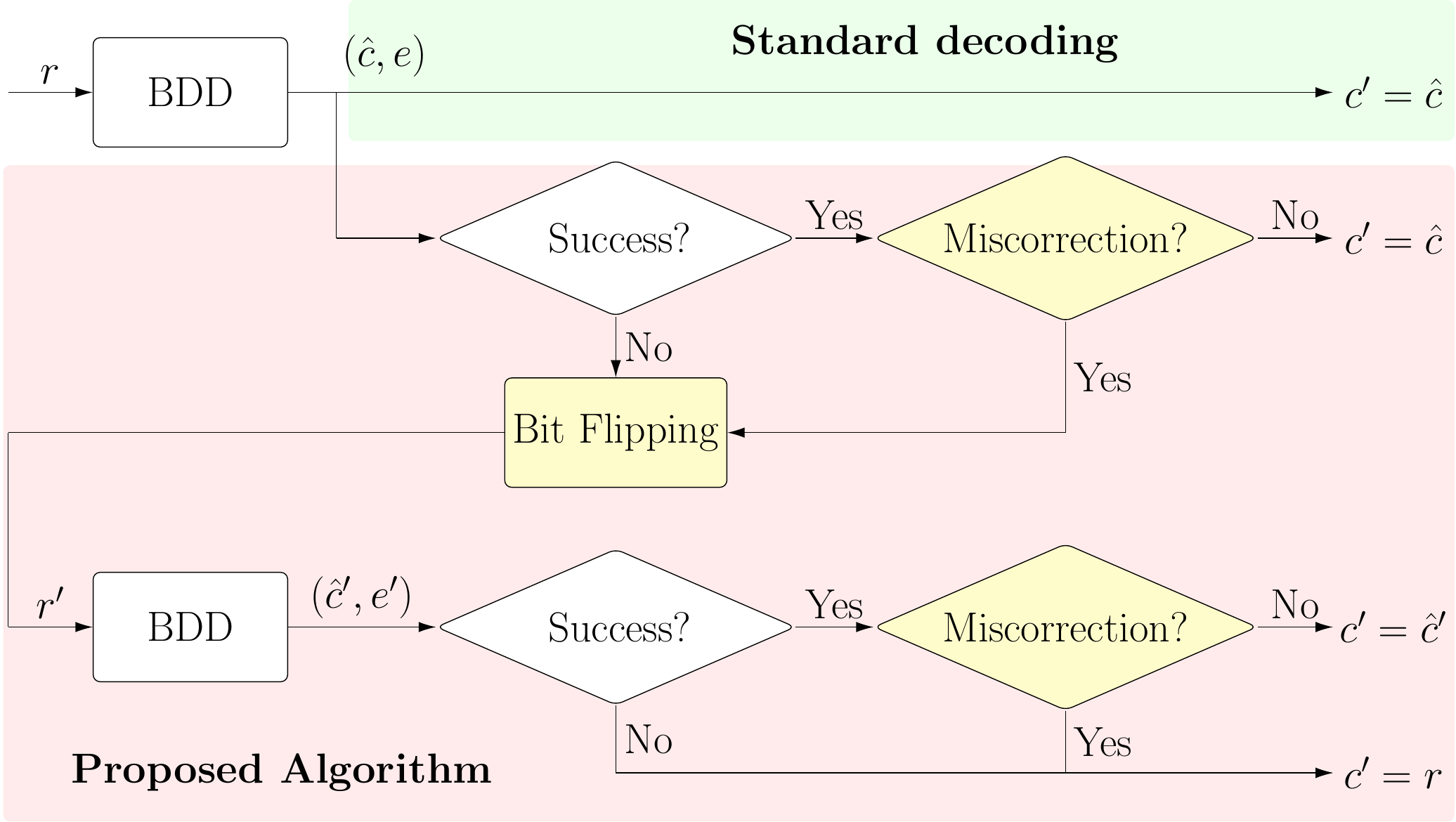}
\caption{Schematic diagram of the proposed algorithm: $r$ is the received sequence
and $c'$ is the output of the staircase decoder. The highlighted blocks use marked LLRs for operation. BDD returns a decoded codeword $\hat{c}$ based on \eqref{BDDequation} and an error pattern $e$.}
\vspace{-2.3ex}
\label{fig: Method}
\end{figure}

The schematic diagram of the proposed algorithm is shown in Fig. \ref{fig: Method} (red area). Compared to standard SCC decoding (green area), which always accepts the decoding result $\hat{c}$ of BDD, the proposed algorithm further checks the decoding status of BDD. If BDD successfully decodes $r$, miscorrection detection is performed. Furthermore, bit flipping (BF) is proposed as a way to handle decoding failures and miscorrections. In this section, we explain the steps in the proposed algorithm.  

Our  proposed algorithm can in principle be applied to all received sequences $r$ within $L$ SCC blocks. However, due to the iterative sliding window decoding structure applied to SCCs, most of the errors are known to be located in the last two blocks. To keep the complexity and latency low, we will therefore only use our algorithm on the received sequences from the last two blocks of the window. Therefore, from now on we only consider rows of the matrix $[\boldsymbol{Y}^{T}_{i+L-2} \boldsymbol{Y}_{i+L-1}]$.


\subsection{Decoding Success: Improved Miscorrection Detection}

To avoid miscorrections, it was suggested in \cite{SmithPhD} to reject the decoding result of BDD applied to $[\boldsymbol{Y}^{T}_{i+L-2} \boldsymbol{Y}_{i+L-1}]$ if the decoded codeword would cause conflicts with zero-syndrome codewords in $[\boldsymbol{Y}^{T}_{i+L-3} \boldsymbol{Y}_{i+L-2}]$. This method protects bits in $\boldsymbol{Y}_{i+L-2}$ but cannot handle bits in the last block $\boldsymbol{Y}_{i+L-1}$. We propose to enhance this method by using marked bits in $\boldsymbol{Y}_{i+L-1}$. In particular, we add one additional constraint to the algorithm in \cite{SmithPhD}: no HRBs in $\boldsymbol{Y}_{i+L-1}$ shall ever be flipped.

The reliability of a bit is given by the absolute value of its LLR, a high value indicating a more reliable bit. Therefore, a threshold $\delta$ is set to decide if the bit is HR. If $|\lambda_{l,k}| \geq \delta$, the corresponding bit is marked as an HRB. The decision of the staircase decoder will therefore be marked as a miscorrection if the decoded codeword causes conflicts with zero-syndrome codewords in $[\boldsymbol{Y}^{T}_{i+L-3} \boldsymbol{Y}_{i+L-2}]$, \emph{or} if the decoded codeword flips a bit whose LLR satisfies $|\lambda_{l,k}| \geq \delta$. 

\begin{example}\label{Example.1}
Fig. \ref{fig: SCC-example} shows a decoding window with $w=6$ and $L=5$ and a component code $\mathcal{C}$ with $t=2$ ($d_{0}=6$). Following the notation of \cite{Christian1}, a pair $(i,j)$ is used to specify the location of a component codeword in each window, where $i\in \{1,2,\ldots,L-1\}$ indicates the position relative to the current window and $j \in \{1,2,\ldots, w\}$ indicates the corresponding row or column index in the matrix of two neighbor blocks. A triple $(i,j,k)$ is used to indicate the $k$th bit in the component codeword $(i,j)$, where $k \in \{1,2,\ldots,2w\}$. For example, the component codewords $(1,2)$ and $(3,1)$ are highlighted with light magenta, while bits $(1,2,11)$ and $(3,1,4)$ are highlighted with dark magenta. The bit sequence $(3,1)$ is a codeword in $[\boldsymbol{Y}^{T}_{i+2} \boldsymbol{Y}_{i+3}]$ whose syndrome is equal to zero. The cells filled with dark yellow are the ones marked as HRBs. 

\begin{figure}[t]
\centering
\includegraphics[width=0.4\textwidth]{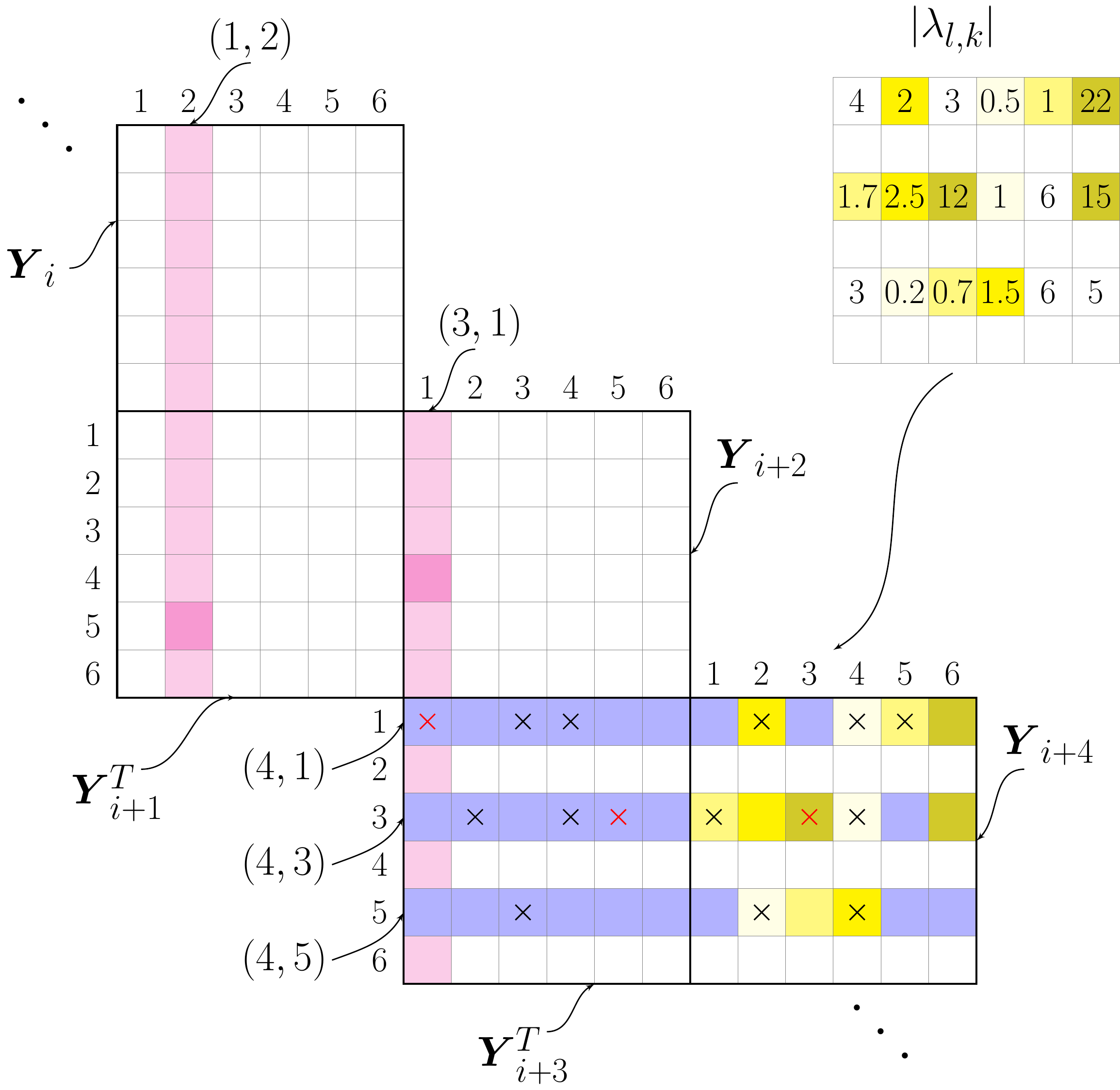}
\caption {Decoding example ($w=6$, $L=5$, $t=2$): black crosses are received errors after channel transmission and red crosses indicate miscorrections after BDD. Dark yellow cells are marked HRBs, while light yellow cells are marked HUBs. Lighter yellow color indicates a smaller value of $|\lambda_{l,k}|$.}
\vspace{-2.3ex}
\label{fig: SCC-example}
\end{figure}

After transmission, the received bit sequences for $(4,1)$ and $(4,3)$ have $5$ and $4$ errors (black crosses), resp. When applying BDD, miscorrections (red crosses) occur. For the received bits in $(4,1)$, BDD detects bit $(4,1,1)$ as an error and suggests to flip it. However, because it is involved in the zero-syndrome codeword $(3,1),$ it will be identified as a miscorrection by both our MD algorithm and the one in \cite{SmithPhD}. For the received bits in $(4,3)$, however, the suggested flipping bit $(4,3,5)$ in $\boldsymbol{Y}_{i+L-2}$ is not involved in any zero-syndrome codewords, and thus, \cite{SmithPhD} would fail to detect this miscorrection. The bit $(4,3,9)$ is a HRB, and thus, our MD algorithm will successfully identify it as a miscorrection.
\end{example}

The MD algorithm in \cite{SmithPhD} does not always detect the miscorrections. The new rule we introduced (never flip HRBs in $\boldsymbol{Y}_{i+L-1}$) is only heuristic and does not guarantee perfect MD either. For example, our MD algorithm fails when no bits are flipped by BDD because $r=\tilde{c}\in\mathcal{C}$. Nevertheless, as we will see later, our MD algorithm combined with bit flipping (see next Sec.) gives remarkably good results with very small complexity increase.

\subsection{Decoding Failures and Miscorrections: Bit Flipping}




To deal with decoding failures and miscorrections, we propose to flip bits (see BF block in Fig.~\ref{fig: Method}). The main idea is to flip certain bits in $r$ and make the resulting sequence $r'$ (after BF) closer to $c$ in Hamming space. In particular, the proposed BF aims at making the Hamming distance between $r'$ and $c$ equal to $t$ so that BDD can correct $r'$ to the transmitted codeword $c$. Two cases are considered by our proposed algorithm: (1) decoding failures, and (2) miscorrections.

\textbf{Case 1 (Decoding Failures):} We target received sequences with $t+1$ errors. In this case, we flip a HUB with the lowest absolute LLR. The intuition here is that this marked bit was indeed one flipped by the channel. In the cases where the marked bit corresponds to a channel error, the error correction capability of the code $\mathcal{C}$ is effectively increased by $1$ bit.

\textbf{Case 2 (Miscorrections):} We target miscorrections where BDD chooses a codeword $\tilde{c}\in\mathcal{C}$ at MHD of $c$. The intuition here is that most of the miscorrections caused by BDD will result in codewords at MHD from the transmitted codeword. When a miscorrection has been detected, our algorithm calculates the number of errors detected by BDD. This is equal to $d_{\textrm{H}}(r,\tilde{c})= w_{\textrm{H}}(e)$. Then, our algorithm flips $d_{0}-w_{\textrm{H}}(e)-t$ bits, which in \emph{some cases} will result in $r'$ that satisfies $d_{\textrm{H}}(c,r')=t$. This will lead BDD to find the correct codeword. More details are given in Examples~\ref{Example.2} and~\ref{Example.3}.
Again using the intuition that bits with the lowest reliability are the most likely channel errors, our BF algorithm flips the most unreliable $d_{0}-w_{\textrm{H}}(e)-t$ bits. In practices, this means that out of $n_c$ code bits per codeword, only $d_{0}-w_{\textrm{H}}(e)-t < t+1$ (or $t+2$ for extended BCH codes) HUBs need to be marked (and sorted). The BF block (see Fig.~\ref{fig: Method}) chooses the number of marked bits to flip based on this sorted list and the Hamming weight of the error pattern. 

\begin{figure}[t]
\centering
\includegraphics[width=0.38\textwidth]{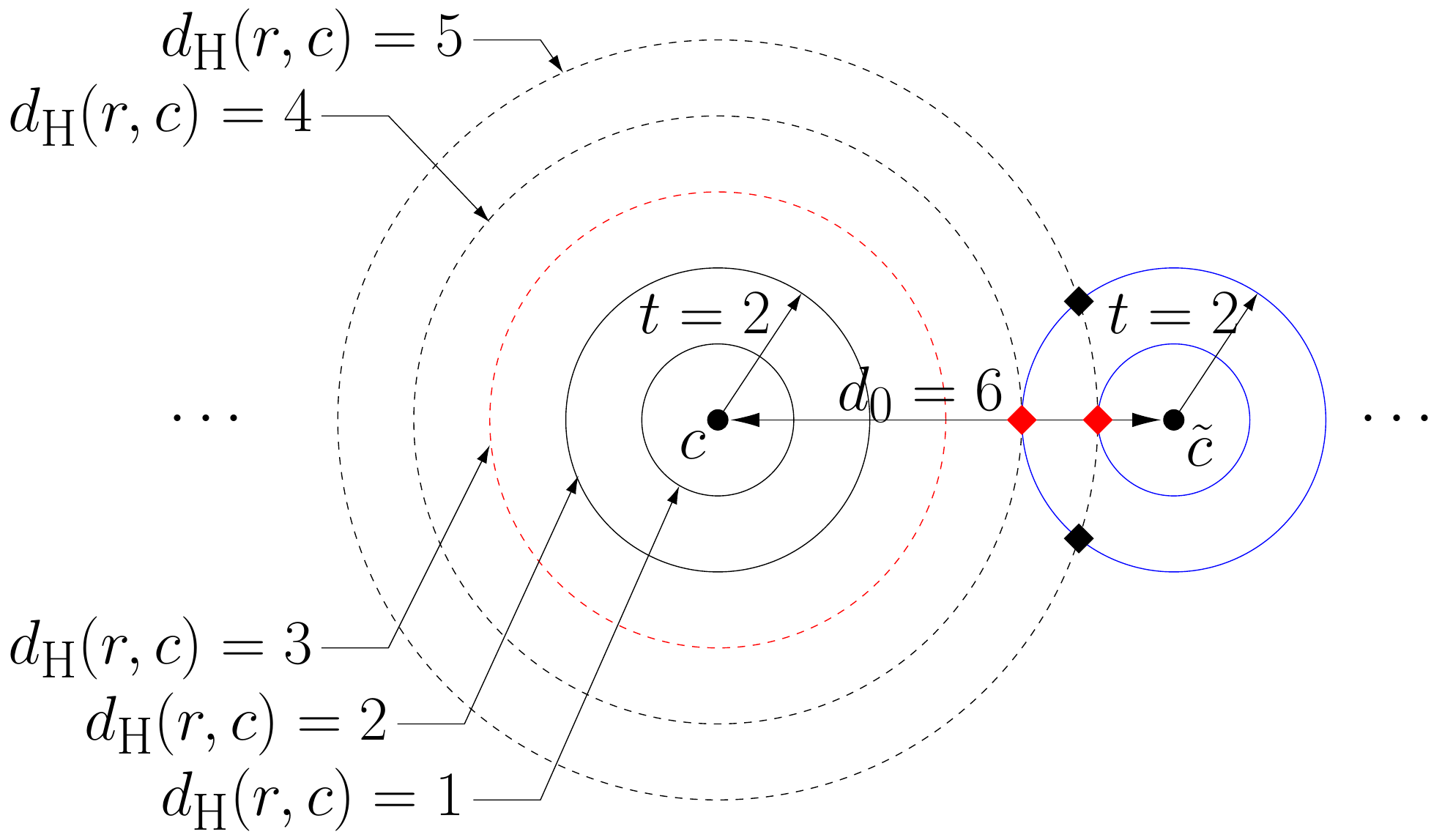}
\caption {Schematic representation of BDD: $c$ is the transmitted codeword and $\tilde{c}\in\mathcal{C}$ is another codeword at MHD $d_{0}=6$. 
}
\vspace{-2.2ex}
\label{fig: BDD-example}
\end{figure}

\begin{example}\label{Example.2}
Fig. \ref{fig: BDD-example} shows a representation of BDD ($t=2$ and $d_{0}=6$), where the black dots represent the transmitted codeword $c$ and another codeword $\tilde{c}\in\mathcal{C}$ with $d_{\textrm{H}}(c,\tilde{c})=d_{0}$. The red dashed circle and solid blue circles correspond to locations of $r$ for Cases 1 and 2, resp. The bit sequence $(4,5)$ in Fig.~\ref{fig: SCC-example} ($3$ errors) would lie on the red dashed circle, while sequences $(4,1)$ and $(4,3)$ correspond to red diamonds ($5$ and $4$ errors, resp.). For the latter two bit sequences, provided that we flip the correct bits (flipping $3$ and $2$ marked bits, resp.), will give an $r'$ with $d_{\textrm{H}}(c,r')=t$ which can be correctly decoded. 
Fig.~\ref{fig: SCC-example} also shows two other miscorrections (black diamonds) which our BF algorithm cannot deal with. In both cases, $w_{\textrm{H}}(e)=2$, and thus, flipping $d_{0}-w_{\textrm{H}}(e)-t=2$ marked bits will not bring $r$ close enough to $c$.
\end{example}

\begin{example}\label{Example.3}
Light yellow cells in Fig. \ref{fig: SCC-example} show the marked 3 HUBs with the lowest reliability within that codeword. The lighter yellow color indicates a smaller value of $|\lambda_{l,k}|$. In this example, BDD fails to decode bit sequence $(4,5)$. Fortunately, $(4,5,8)$ corresponds to the marked HUB with smallest $|\lambda_{l,k}|$. Thus, it will be flipped after BF, and then the remaining 2 errors $(4,5,3)$ and $(4,5,10)$ will be fully corrected by applying BDD again. This corresponds to Case 1.

For bit sequences (4,1) and (4,3), the decoding results of BDD are identified as miscorrections (as explained in Example \ref{Example.1}) with $w_{\textrm{H}}(e)=1$ and $w_{\textrm{H}}(e)=2$, resp. According to the BF rule for miscorrections, 3 and 2 bits with smallest $|\lambda_{l,k}|$ among the marked HUBs, i.e., (4,1,8), (4,1,10), (4,1,11) in (4,1), and (4,3,7), (4,3,10) in (4,3), will all be flipped. As a result, only 2 errors are left in (4,1) and (4,3), which are within the error correcting capability of BDD. This corresponds to Case 2.
\end{example}

BF will not always result in the correct decision. As shown in Example~\ref{Example.2}, this is the case for certain miscorrections (black diamonds in Fig.~\ref{fig: SCC-example}). Additionally, miscorrections for codewords at distances larger than $d_0$ are not considered either. Finally, marked LLRs might not correspond to channel errors. In all these cases, either decoding failures or miscorrections will happen. To avoid these cases, our proposed algorithm includes two final checks after BF and BDD (see lowest part of Fig.~\ref{fig: Method}): successful decoding and MD.


\section{Simulation Results}


The component codes used for simulations are extended BCH codes with $t=2$. The decoding window size is $L=9$, and the maximum  number of iterations is $7$. The LLR threshold $\delta$ in the MD algorithm is set to $10$, which gives the best performance for $R=0.87$ and 2-PAM. Optimization of $\delta$ for different SNRs could provide additional gains.

\begin{figure}[t]
\centering
\includegraphics[width=0.5\textwidth]{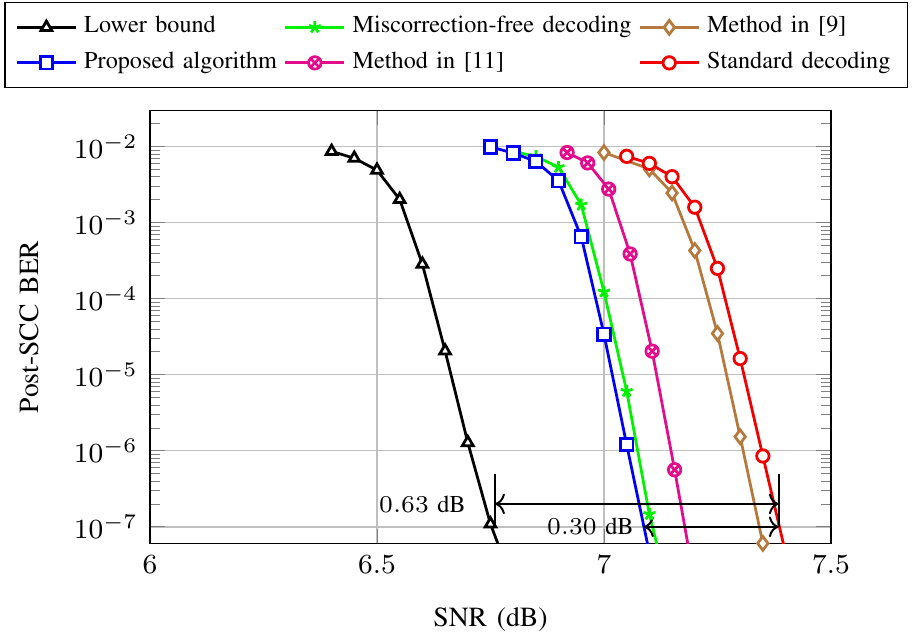}
\caption{Post-SCC BER vs. SNR for code rate $R=0.87$ and and $2$-PAM.}
\vspace{-2ex}
\label{fig: SCC for m=8,t=2}
\end{figure}

\begin{figure*}[t]
\centerline{\includegraphics[width=1\textwidth]{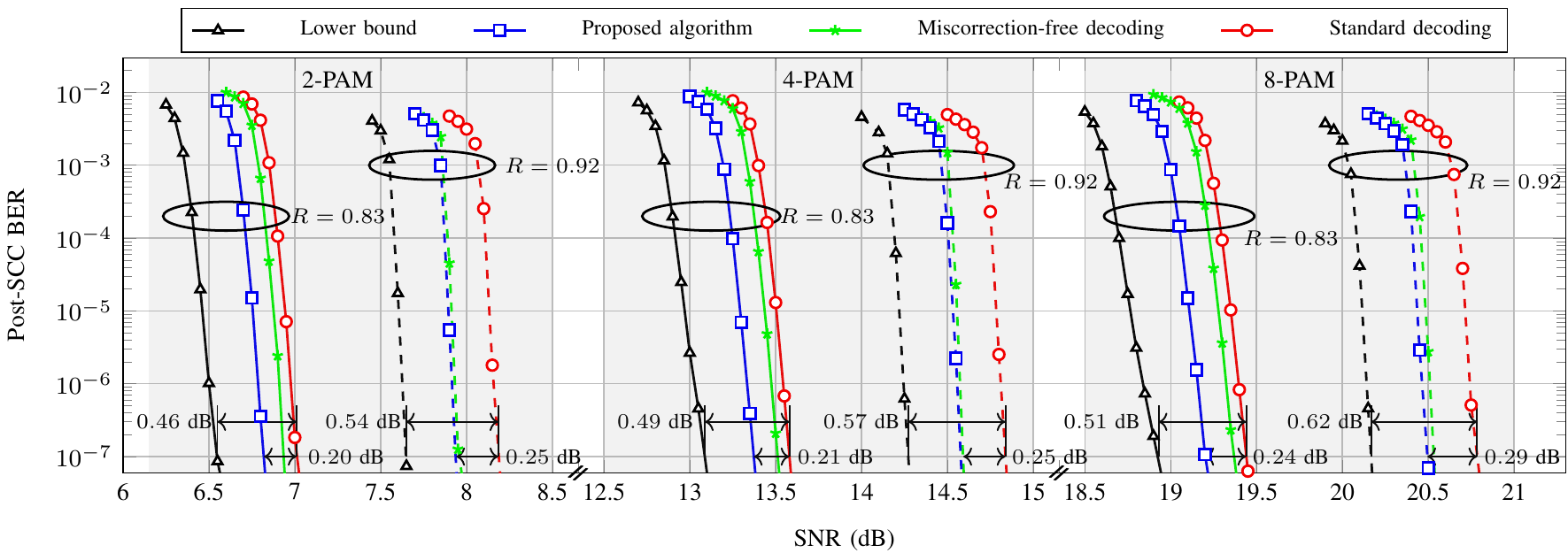}}
\caption{Post-SCC BER vs. SNR for code rates $R=0.83$ and $R=0.92$ with 2-PAM, 4-PAM, and 8-PAM modulation formats.}
\vspace{-1ex}
\label{fig: SCC for m=9,t=2}
\end{figure*}

\begin{example}
Consider $2$-PAM and a SCC with $R=0.87$, i.e., with BCH codes $(256,239,2)$ ($w=128$) as component code. These parameters are chosen to compare our algorithm with results presented in the literature. These results are shown in Fig. \ref{fig: SCC for m=8,t=2}. Two baselines are: standard decoding where miscorrections are not dealt with (circles), and miscorrection-free decoding (stars). The latter is obtained via a genie BDD decoder which corrects the received sequence only when the number of errors is not more than $t$. This figure also shows the performance of previously proposed methods: \cite{SmithPhD} and \cite{Christian1}. 

The performance of our algorithm (squares) is also shown in Fig. \ref{fig: SCC for m=8,t=2}, which is $0.3$~dB better than standard decoding. Our algorithm also outperforms both \cite{SmithPhD} and \cite{Christian1}. These two methods only prevent miscorrections, so their performance is bounded by the miscorrection-free case. Although our algorithm only deals with miscorrections related to the last block of each window, it outperforms the miscorrection-free case. This is due to its additional ability to better deal with miscorrections and decode even when BDD initially fails.

Fig. \ref{fig: SCC for m=8,t=2} also shows a lower bound for our proposed algorithm (triangles). This bound is obtained by a genie decoder which emulates a best-case scenario for our algorithm. This genie decoder is assumed to be able to ideally identify all miscorrections in the last two blocks of the window. This corresponds to having an idealized MD block in the top part of Fig.~\ref{fig: Method}. 
The genie decoder also emulates an idealized assumption on what the BF block in Fig.~\ref{fig: Method} can do. For this, we assume that the decoder knows exactly which bits in the last two blocks are errors. If a given sequence has $t+j$ errors ($j=1$ for Case 1, or $j=d_{0}-w_{\textrm{H}}(e)-t$ for Case 2), and at least $j$ errors are located in the last block, the genie decoder flips $j$ errors in the last block, and then the received sequence is correctly decoded. If less than $j$ errors are located in the last block, the genie decoder declares a failure. The results show that the maximum potential gain for our receiver structure (for $2$-PAM, $R=0.87$, and $t=2$) is $0.63$~dB. Our algorithm achieved half of this gain with very small added complexity.
\end{example}



\begin{example}
Fig.~\ref{fig: SCC for m=9,t=2} shows the simulation results of the proposed algorithm for 2-PAM, 4-PAM and 8-PAM. For each modulation format, two code rates are considered: $R=0.83$ and $R=0.92$ with code parameters $(228,209,2)$ and $(504,485,2)$, resp. These parameters are obtained by shortening the extended BCH code $(512,493,2)$ by $284$ and $8$ bits, resp. It can be seen from Fig.~\ref{fig: SCC for m=9,t=2} that for different modulation formats and code rates, our proposed algorithm always outperforms the miscorrection-free case. When compared to standard staircase  decoding, the achieved gains are between $0.20$~dB and $0.29$~dB, while the obtained maximum potential gains are between $0.46$~dB and $0.62$~dB. The results also show that the gains increase as the modulation size increases.
\end{example}

\section{Conclusions}
In this paper, a novel decoding algorithm for staircase codes was proposed. This algorithm is based on simple modification of the standard hard-decision-based staircase decoder and relies on the idea of marking bits. The algorithm consists of an improved miscorrection-detection mechanism and a bit-flipping operation to effectively increase the error correcting performance of bounded-distance decoding. Large gains compared to standard SCC decoding were obtained. A precise complexity evaluation of the algorithm is left for further investigation.

\medskip

\bibliographystyle{IEEEtran}
\bibliography{IEEEabrv,refs_ISTC}
\end{document}